\documentstyle[12pt,epsf,psfig]{article}
\def\J{$J/\psi$}

\def\e{\epsilon}

\def\N{$n_{\rm cl}$}
\def\n{n_{\rm cl}}
\def\s{s_{\rm cl}}
\def\S{s_{\rm cl}}

\def\be{\begin{equation}}
\def\ee{\end{equation}}

\def\lsim{\raise0.3ex\hbox{$<$\kern-0.75em\raise-1.1ex\hbox{$\sim$}}}
\def\gsim{\raise0.3ex\hbox{$>$\kern-0.75em\raise-1.1ex\hbox{$\sim$}}}

\def\NP{{ Nucl.\ Phys.\ }}
\def\PL{{ Phys.\ Lett.\ }}
\def\PR{{ Phys.\ Rev.\ }}

\def\PRL{{ Phys.\ Rev.\ Lett.\ }}

\def\ZP{{ Z.\ Phys.\ }}

\begin{document}

\noindent May 21, 1998 \hfill BI-TP 98/11

\vskip 1.5cm

\centerline{\Large{\bf ~Deconfinement and Percolation}\footnote{Talk
given at the International Symposium {\sl QCD at Finite Baryon Density},
April 27 - 30, 1998, Bielefeld, Germany; {\sl Proceedings} to appear in
Nuclear Physics A.}}

\vskip 1.0cm

\centerline{\bf Helmut Satz}

\bigskip

\centerline{Fakult\"at f\"ur Physik, Universit\"at Bielefeld}

\par

\centerline{D-33501 Bielefeld, Germany}

\vskip 1.0cm

\centerline{\bf Abstract:}

Using percolation theory, we derive a conceptual definition of
deconfinement in terms of cluster formation. The result is readily
applicable to infinite volume equilibrium matter as well as to finite
size pre-equilibrium systems in nuclear collisions.

\vskip 1.0 cm

\noindent
{\bf 1.\ Introduction}

\medskip

Quantum chromodynamics predicts that strongly interacting matter will
undergo a transition from a low density medium consisting of
colour-neutral hadrons to a plasma of deconfined coloured
quarks and gluons at high densities. This transition has been confirmed
and studied in detail in finite temperature lattice QCD at zero
baryon number density; nevertheless, even there it can so far be
formulated concisely only in the limits of infinite and of vanishing
quark mass.

\medskip

Infinite quark mass leads to pure colour $SU(N)$ gauge theory. Here the
Polyakov loop $L(T)$ provides an order parameter suitable to define the
confinement-deconfinement phase transition, driven by the underlying
center $Z_N \!\subset\! SU(N)$ symmetry of the Lagrangian. Confined
states share this symmetry, but it is spontaneously broken in the
deconfined phase. For QCD with quarks of mass $m_q < \infty$, the
symmetry is broken explicitly in the Lagrangian.

\medskip

In the limit of vanishing quark mass, the Lagrangian becomes invariant
under chiral transformations. Now the quark condensate $\langle \psi
{\bar \psi} \rangle(T)$ constitutes an order parameter, distinguishing
the hadronic phase of spontaneously broken chiral symmetry from the
quark-gluon plasma, in which this symmetry is restored. In other words,
in the confined phase the massless quarks `dress' themselves to become
massive constituent quarks, while in the deconfined phase the massless
current quarks of the Lagrangian are recovered. Here as well, the
introduction of a finite quark mass $m_q$ explictly breaks the
relevant symmetry of the Lagrangian.

\medskip

The real world lies between these two limits, and there the nature of
the transition is less clear (Fig.\ 1). The aim of the present paper
is to study this region of broken symmetries and to look for a more
general description of deconfinement which would be applicable there
as well. Conceptually, the transition from
hadronic matter to quark-gluon plasma seems rather transparent, no
matter what the quark mass is. Once the density of constituents becomes
so high that several hadrons have considerable overlap, it no longer
makes sense to partition the underlying quarks into colour-neutral
bound states. Instead, there appear clusters much larger than hadrons,
within which colour is not confined. This suggests that deconfinement
is related to cluster formation, and since that is the central topic of
percolation theory, possible connections between percolation and
deconfinement were discussed already quite some time ago on a rather
qualitative level \cite{Baym,Celik}. In the meantime, however, the
interrelation of geometric cluster percolation and critical behaviour
of thermal systems has become much better understood \cite{S-A}, and we
want to apply this understanding to clarify the nature of deconfinement.

\begin{figure}[h]
\vspace*{-0mm}
\centerline{\psfig{file=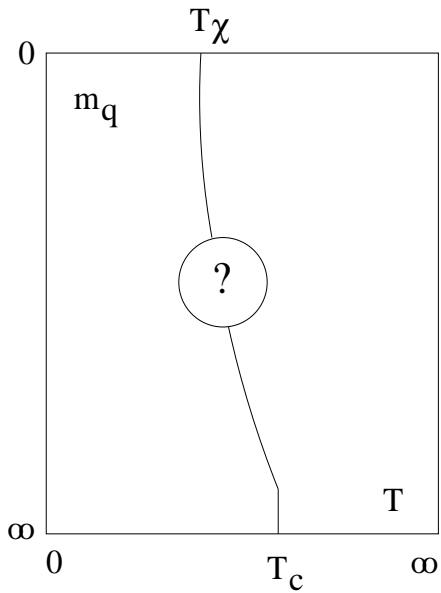,height= 80mm,angle= -90}}
\caption{Critical behaviour in finite temperature QCD.}
\end{figure}

\medskip

In section 2, we shall recall some aspects concerning the relation
between deconfinement and chiral symmetry restoration. The main results
of percolation theory for lattice systems will be summarized in section
3 and extended to continuum percolation in section 4. This will then
provide the basis for the study of percolation in finite temperature
and finite density QCD in section 5. Finally, in section 6, we shall
consider applications to deconfinement in nuclear collisions.

\bigskip

\noindent
{\bf 2.\ Critical Behaviour in Finite Temperature QCD}

\medskip

We want to consider here in particular the role of the `bare' quark mass
$m_q$ in the Lagrangian
\be
{\cal L}~=~-{1\over 4}F^a_{\mu\nu}F^{\mu\nu}_a~-~{\bar\psi}_\alpha
(i \gamma^{\mu}\partial_{\mu}-g \gamma^{\mu}A_{\mu})^{\alpha\beta}
\psi_\beta~
-m_q~{\bar\psi}_\alpha \psi_\alpha
\label{2.1}
\ee
for the critical behaviour of statistical QCD. From the point of view
of pure gauge theory, the introduction of $m_q$ breaks explicitly the
center $Z_N$ symmetry of the Lagrangian and thus is
like turning on an external magnetic field $H \sim 1/m_q$ in the
corresponding $Z_N$ spin theory; for infinite quark mass, we recover
the thermodynamics of pure gauge theory, just as $H=0$ leads back to
the $Z_N$-symmetric spin system. For sufficiently small quark masses,
one might thus expect a smooth temperature variation of the Polyakov
loop \cite{Hasenfratz}, similar to the magnetization pattern of
spins in a strong external field. However, such a behaviour is
not observed; even for quark masses approaching the chiral limit,
the Polyakov loop
\be
L(T,m_q) \sim \exp\{-V(m_q)/T\} \label{2.2}
\ee
varies very rapidly at those `critical' temperatures at which energy
density and chiral condensate vary strongly; here $V(m_q)$ denotes the
interquark potential in the limit of large separation. The reason
for this persistence of an `almost' critical behaviour of $L(T,m_q)$ in
$T$ seems evident: the spontaneous breaking of the chiral symmetry for
$m_q=0$ produces an effective `constituent' quark mass $M_q \simeq
M_h/2$, equal to about half the mass $M_h$ of the basic (non-Goldstone)
meson. For $m_q \not=0$, chiral symmetry is explicitly broken and
$M_q(m_q)$ thus varies with $m_q$ in the form shown in Fig.\ 2.
Equivalently, the potential string breaks in the large distance limit
not for $2m_q$, but rather for $M_h \simeq 2M_q$, so that even for
$m_q \to 0$, a finite energy is still required for string breaking.
It is $M_q^{-1}$, not the inverse bare quark mass $m_q^{-1}$, which
acts as external field, and hence this field always remains quite weak.
The range of variation of $L(T,m_q)$ thus lies between the singular
pure gauge form $L(T,m_q\!=\!\infty)$ and a limiting form
$L(T,m_q\!=\!0)$ which still varies strongly with $T$ (Fig.\ 3).

\begin{figure}[h]
\vspace*{-0mm}
\centerline{\psfig{file=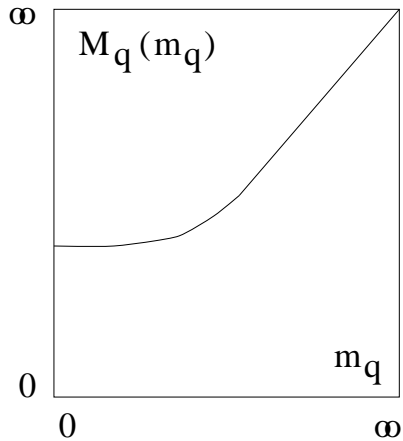,height= 60mm,angle= -90}}
\caption{Variation of the constituent quark mass with bare quark mass.}
\end{figure}

\medskip

\begin{figure}[h]
\vspace*{-0mm}
\centerline{\psfig{file=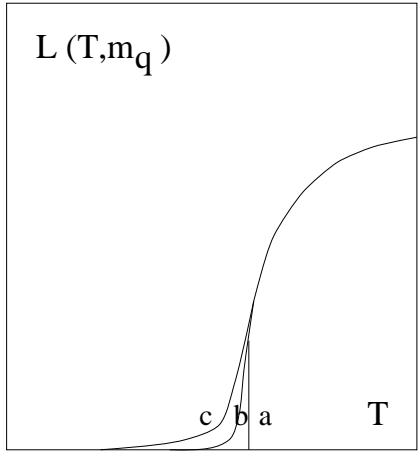,height= 60mm,angle= -90}}
\caption{The temperature dependence of the Polyakov loop (a) for pure gauge
theory ($m_q \to \infty$), (b) for QCD with light quarks
($0 < m_q < \infty$), and (c) in the chiral limit ($m_q=0$).}
\end{figure}

The thermal pattern of $L(T,m_q)$ should be contrasted to that of the
chiral condensate  $\langle \psi {\bar \psi} \rangle (T,m_q)$. For
$m_q=0$, the spontaneous breaking of the chiral symmetry at low
temperatures disappears when $T=T_{\chi}$, and from
there on chiral symmetry is restored. The introduction of a
non-vanishing bare quark mass $m_q$ explicitly breaks chiral symmetry,
so that here $m_q$ (and not its inverse) acts like an external field in
the corresponding $O(4)$ spin system \cite{P-W}; only for
$m_q = 0$ is there genuine spontaneous symmetry breaking, and when $m_q$
becomes large, the temperature variation of $\langle \psi {\bar \psi}
\rangle (T,m_q)$ is washed out completely.

\medskip

The `transition' line between $m_q=\infty$ and $m_q=0$ in Fig.\ 1 can
thus be characterized as follows: the singular behaviour of
$L(m_q=\infty)$ changes into an `almost' singular behaviour as the quark
mass becomes finite, with $L(M_q[m_q=0])$ as the chiral limit. Any
(spontaneous or explicit) breaking of chiral symmetry assures that $M_q
\not=0$ and hence limits the strength of the effective external field.
The associated `almost' singular behaviour of thermal observables such
as entropy or energy density is thus related to deconfinement as far as
the overall variation is concerned; the chiral features determine how
`smooth' this variation can become, and explicit or spontaneous chiral
symmetry breaking keeps it always very pronounced. -- While these
considerations provide a reason why the deconfinement transition, as
defined by the behaviour of the Polyakov loop, `almost' persists in
the chiral limit, they do not explain why deconfinement and chiral
symmetry restoration effectively coincide at zero baryon number
density; all arguments appear to remain valid if chiral symmetry were
restored above deconfinement.

\medskip

We now turn to percolation theory, with the aim of obtaining a
characterization of deconfinement which remains valid for all values of
the quark mass $m_q$.

\bigskip

\noindent
{\bf 3.\ Percolation and the Ising Model}

\medskip

For simplicity, consider a two-dimensional square lattice of linear
size $L$; we randomly place identical objects on $N$ of the $L^2$
lattice sites. With increasing $N$, adjacent occupied sites will begin
to form growing clusters or islands the sea of empty sites. Define
$n_p$ to be the lowest value of the density $n = N/L^2$ for which the
origin belongs on the average (i.e., calculated by averaging over many
randomly generated configurations) to a cluster reaching the edge of
the lattice. In the limit of infinite lattice size, we then have
\be
P(n) \sim \left(1 - {n_p \over n}\right)^{\beta_p},~~~n \geq n_p,
\label{3.1}
\ee
for $n$ approaching $n_p$ from above, where $P(n)$ denotes the probability
that the origin belongs to an infinite
cluster. Since $P(n)=0$ for all $n \leq n_p$ and non-zero for all $n
> n_p$, it constitutes an order parameter for percolation:
$\beta_p=5/36$ is the critical exponent which governs the vanishing of
$P(n)$ at $n=n_p$ in two dimensions; in three dimensions, it becomes
$\beta=0.41$ \cite{Isi}.

\medskip

For $n < n_p$, a quantity of particular interest is the
`mean' cluster size $S(n)$, defined as the average size (the number of
connected occupied sites) of a cluster containing the origin of the
lattice. As $n$ approaches $n_p$ from below, $S(n)$ diverges at the
percolation point as
\be
\S(n) \sim (n_p - n)^{-\gamma_p},~~~n\leq n_p \label{3.2}
\ee
with $\gamma_p=43/18$ (1.80) as the $d=2$ (3) critical exponent for the
divergence.

\medskip

Instead of the site percolation just described, a random placement of
bonds on the lines between adjacent sites leads to bond percolation,
with the same critical exponents. Clusters now consist of regions of
connected bond lines. While the critical behaviour is universal, the
value of the percolation threshold $n_p$ depends on the type of
percolation considered; on a two-dimensional square lattice,
it is 0.5 for bond and 0.59 for site percolation \cite{Isi}.

\medskip

We now turn to the Ising model, defined by the Hamiltonian
\be
{\cal H}_I = -J \sum_{i,j} s_i s_j ~-~ H \sum_{i} s_i, ~s_i=\pm 1,
\label{3.3}
\ee
where the first sum runs only over nearest neighours on the lattice,
with $J$ denoting the exchange energy between spins and $H$ an external
field. For $H=0$, the Hamiltonian (3) has a global $Z_2$
invariance ($s_i \to -s_i~ \forall~ i$), and the magnetization $m =
\langle s \rangle$ probes whether this invariance is spontaneously
broken. As is well-known, such spontaneous symmetry breaking occurs
below the Curie point $T_c$, with
\be
m(T,H=0) \sim \left( 1 - {T_c \over T} \right)^{\beta_m} \label{3.4}
\ee
governing the vanishing of $m(T,H=0)$ as $T \to T_c$ from below.
The well-known Onsager solution gives with $\beta_m=0.125$ for $d=2$
a 10 \% smaller value than the $\beta_p=5/36 \simeq 0.139$ found for the
percolation exponent.

\medskip

Since the Ising model also produces clusters on the lattice, consisting
of connected regions of aligned up or down spins, the relation between
its thermal critical behaviour at $T_c$ and the onset of geometric
percolation is an obvious question which has been studied extensively
in recent years. Two important distinctions have emerged: the nature of
what is defined as a cluster \cite{C-K}, and the behaviour of clusters
in a non-vanishing external field \cite{Kertesz,S-W}.

\medskip

The geometric clusters considered in percolation theory consist simply
of connected regions of spins pointing in the same direction. In
the Ising model, there is a thermal correlation between spins on
different sites; this vanishes for $T \to \infty$. Correlated regions
in the Ising model (we follow the usual notation and call them
`droplets', to distinguish them from geometric clusters) thus disappear
in the high temperature limit. In contrast, the
geometric clusters never vanish, since the probability for a finite
number of adjacent aligned spins always remains finite; it increases
with dimension because the number of neighbours does. Hence from the
point of view of percolation, there are more and bigger clusters than
there are Ising droplets. As an immediate consequence, percolation
occurs in general before the onset of spontaneous magnetization.

\medskip

If percolation is to provide the given thermal critical behaviour, the
definition of cluster has to be changed such that the modified
percolation clusters coincide with the correlated Ising droplets
\cite{C-K}. This is achieved by assigning to some pairs of adjacent
aligned spins in a geometric cluster an additional bond correlation,
randomly distributed with the density
\be
n_b = 1 - \e^{-2J/kT}, \label{3.5}
\ee
where $[-2J/kT]$ just corresponds to the energy required for flipping
an aligned into an opposing spin. The modified percolation clusters
then consist of aligned spins which are bond connected. Only for $T=0$
are all aligned spins bonded; for $T >0$, some aligned spins in a
purely geometric cluster are not bonded and hence do not belong to the
modified cluster or droplet. This effectively reduces the size of a
given geometric cluster or even cuts it into several modified clusters.
For $T \to \infty$, $n_b \to 0$, so that the geometric clusters still
in existence there are not counted as droplets, solving the problem
mentioned in the previous paragraph.

\medskip

With such a superposition of site and bond percolation (`s/b'), full
agreement between percolation and thermal critical behaviour of
the Ising model is achieved. An infinite percolation cluster is now
formed for the first time at $T_c$, the cluster size coincides
with that of the correlated regions in the Ising model, and numerical
simulations show that the effective critical exponents for the new
$P_{s/b}(n)$ or $S_{s/b}(n)$ become those of the Ising model.

\medskip

However, for $H \not= 0$, Ising model and percolation theory results do
not coincide, neither for purely geometric nor for site/bond defined
percolation clusters. The Ising model for $H \not= 0$ does not lead to
any singularity as function of $T$ and hence does not show any critical
behaviour \cite{L-Y}; the $Z_2$ symmetry responsible for the onset of
spontaneous magnetization is now always broken and $m(T,H\not=0) \not=
0$ for all $T$. On the other hand, the size of connected regions of
aligned and bonded spins increases with decreasing temperature, and
above some critical temperature it diverges. Hence percolation will
occur for any
value of $H$. At $H=\infty$, all spins are aligned, leaving the
bonds as the relevant variables; the system thus percolates at the
critical density for bond percolation, which is 1/2 in two and 0.249 in
three dimensions, for square and cubic lattices, respectively.
From Eq.\ (\ref{3.4}), this corresponds to temperatures $kT_b/J=2.89$
and 6.99 in two and three dimensions, to be compared to the Curie
temperatures $kT_c/J=2.27$ and 4.51, respectively. The corresponding
values at finite $H$ lie between $T_c$ at $H=0$ and $T_b$ at $H=\infty$
and define the so-called Kert\'esz line \cite{S-A,Kertesz}; see Fig.\
4. What happens at this line in terms of dynamics?

\medskip

\begin{figure}[h]
\vspace*{-0mm}
\centerline{\psfig{file=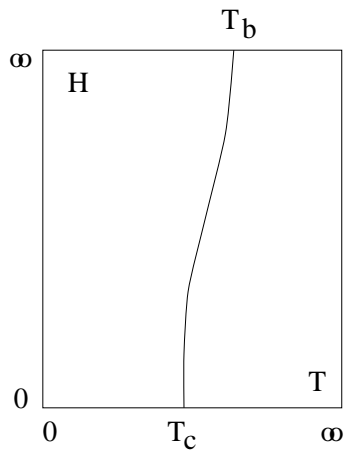,height= 60mm,angle= -90}}
\caption{The K\'ertesz line, defining the percolation temperature in 
the Ising model as function of the external field $H$; $T_c$ denotes
the Curie temperature, $T_b$ the pure bond percolation temperature.}
\end{figure}

Since the Ising model does not lead to singular behaviour in $T$ for
$H\not=0$, in this case even the modified percolation
picture does not agree with the Ising result. The reason for this is
physically quite evident. At sufficiently high temperatures, only finite
size s/b clusters exist, so that the percolation probability vanishes;
at low temperatures, on the other hand, the s/b cluster size diverges,
and hence $P(n)$ remains as genuine order parameter. The existence
of a non-vanishing external field, however, partially aligns in its
direction the spins of the different disconnected finite size clusters
at high temperature \cite{S-W}, so that the overall magnetization never
vanishes. As a result, the critical behaviour due to percolation
persists, while that related to the thermal magnetization pattern
disappears.

\medskip

As is well-known, the Ising model can be reformulated as a lattice
gas, where its critical behaviour describes in a rudimentary way the
vapour-liquid transition. The non-singular behaviour for $H\not=0$
then translates into a continuous cross-over from gas to liquid for
sufficiently high pressure; the two phases are assumed to become
indistinguishable in the relevant pressure-temperature regime. However, the
percolation effect discussed in the previous paragraph seems to contradict
the established credo that `nothing happens above the critical point'. It
is observed \cite{Kertesz,A-S} that for temperatures below the Kert\'esz
line, Ising droplets have a finite surface tension; this vanishes along
the Kert\'esz line, and on its high temperature side, the droplet
energy becomes proportional to their bulk energy. Note that such a
behaviour does not contradict the non-singular thermodynamic behaviour
of the model, since the variable that changes along the Kert\'esz line,
the surface tension of droplets, is not expressible in terms of
thermodynamic observables.

\medskip

In the same vein, we note another illustration of the relation between
percolation and thermal critical behaviour. Consider bond percolation
in the two-dimensional Ising model and imagine that current can flow
between two or more bonded sites. In this case, conductivity sets in at
the percolation point, independent of the thermal critical behaviour of
the Ising model; the system is non-conducting below the percolation
point and conducting above it. This remains true also in s/b
percolation and is independent of whether clusters are partially
aligned by an external field. In other words, we have two independent
critical phenomena, the onset of conductivity and the critical
behaviour of Ising thermodynamics. This suggests a possible framework
for a general characterization of deconfinement. Before addressing this,
however, we want to point out some aspects encountered in the extension
of percolation to continuum systems \cite{Isi}.

\bigskip

\noindent
{\bf 4.\ Continuum Percolation}

\medskip

We start again in two dimensions and consider the distribution of discs
of radius $r$ over some plane area, allowing overlap; as example, we
take a circle of radius $R \gg r$ (see Fig.\ 5). The average cluster
size $\s$ and density \N~(number of discs per cluster) is determined
also here by averaging clusters containing the origin over many randomly
generated configurations. These quantities are
studied as function of the overall density $n$, i.e., the number of
discs per circle $\pi R^2$. In the limit $R \to \infty$, the percolation
probability $P(n)$ and the divergence of the cluster size $S(n)$ are
then given by Eqs.\ (\ref{3.1}) and (\ref{3.2}), with the same critical
exponents; the same remains true for the corresponding
three-dimensional case, with overlapping spheres instead of discs.

\begin{figure}[h]
\vspace*{-0mm}
\centerline{\psfig{file=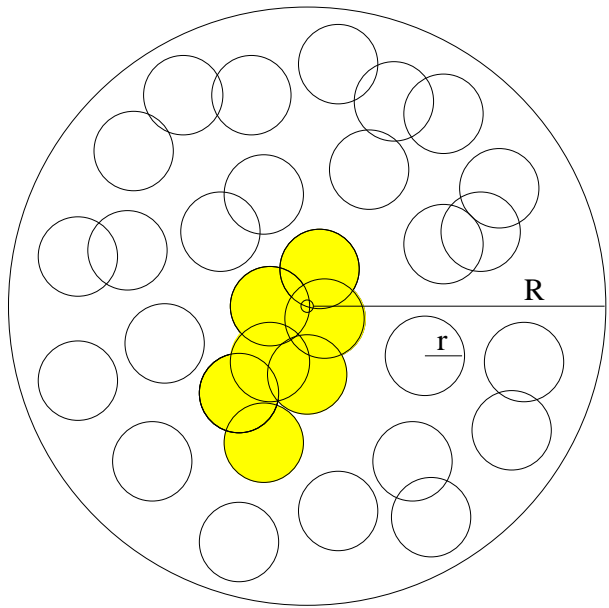,height= 80mm,angle= -90}}
\caption{Two-dimensional continuum clustering.}
\end{figure}

\medskip

In the continuum, the percolation threshold becomes \cite{Isi,Alon}
\be
n_p \simeq {1.18 \over \pi r^2} \label{4.1}
\ee
for $d=2$ and
\be
n_p \simeq {0.34 \over (4 \pi/3) r^3} \label{4.2}
\ee
for $d=3$. In both cases, there is considerable overlap, and at the
percolation point, the fraction of the area (volume) covered by discs
(spheres) is $1 - \exp(-\eta_p)$, where $\eta_p\simeq1.18$ (0.34) for
$d=2$ (3).

\medskip

Besides the discs and spheres considered here, the percolation of
cylinders or ellipsoids will turn out to be of interest to us.
While the critical behaviour at the percolation threshold is universal,
i.e., dependent only on the space dimension of the problem, the
threshold value itself is affected by the geometry of the objects
distributed in space. For ellipsoids, percolation has been studied
numerically in considerable detail \cite{Garboczi}: as representative
illustration, the threshold density for percolation is found to be
\be
n_p \simeq { 0.21 \over (4 \pi/3) a b^2}, \label{4.3}
\ee
in the case of an aspect ratio $a/b=4$ ($a$ is the long, $b$ the short
axis of a prolate ellipsoid).

\medskip

In section 3, we had noted the Coniglio-Klein cluster definition
\cite{C-K}, which resulted in the coincidence of percolation and
thermal critical behaviour of the Ising model. For our later
considerations, the extension of this definition to the continuum is
necessary. This, however, for the moment appears to remain an
interesting open problem in percolation theory \cite{Coniglio}.

\bigskip

\noindent
{\bf 5.\ Percolation in QCD}

\medskip

We now turn to the application of percolation in the study of strongly
interacting matter and consider first pure colour $SU(N)$ gauge theory.
Here the problem is straight-forward: we have to extend the
Coniglio-Klein approach \cite{C-K} to continuum percolation, i.\ e.,
construct a cluster definition which moves the critical behaviour
at the percolation threshold from the universality class of geometric
percolation to that of thermal critical behaviour of a $Z_N$ spin system
\cite{S-Y}. If this is possible, glueball percolation will lead to the
correct deconfinement behaviour as obtained in terms of the Polyakov
loop as order parameter.

\medskip

For the case of QCD with dynamical quarks, it had already been noted in the
first studies\cite{Baym,Celik} that the percolation of spheres of
hadronic size cannot mean deconfinement: using Eq.\ (\ref{4.2}) and a
hadron radius $r = 1.0$ fm, the percolation threshold becomes $n_p
\simeq 0.08$ fm$^{-3}$, i.e., it lies below normal nuclear
density. Hence at this density, the quarks must still remain
effectively localized in colour singlet states, so that the percolation
point here corresponds to the formation of connected nuclear matter.
Deconfinement, it was argued, occurs only when percolation sets in for
some smaller `hard core' of hadrons. The question was how to define
this.

\medskip

In QCD, two fundamental triplet charges are connected by a colour flux
tube or string, with $\sigma \simeq 0.16$ GeV$^2$ as string tension. The
transverse radius $r$ of these strings has the form
\cite{Luescher,Laermann}
\be
r^2 = {1\over \pi \sigma} \ln(l/l_0), \label{5.1}
\ee
where $l_0 \simeq 0.1 - 0.3$ fm is the string formation length
\cite{Alvarez,Schilling}. Extensive lattice QCD studies find that in the
range $0.5 \leq l \leq 2$ fm, $r \simeq 0.2 - 0.3$ fm and varies at most
weakly \cite{Laermann,Schilling}. As a result, a hadron appears as a
cylindrical string of diameter $2r$ and length $l \simeq 1$
fm, oscillating about its center of mass to fill a sphere of that
radius. While the energy density in the string of length $l$ and radius
$r$ is of order 5 GeV/fm$^3$, the oscillation spreads this over a sphere
of radius $l$ and thus reduces the energy density to the 0.2 - 0.3
GeV/fm$^3$ observed for a normal (non-Goldstone) meson.

\medskip

Given such a partonic substructure of a hadron, it becomes clear that
deconfinement requires the percolation of the underlying strings.
Evidently (compare Eqs.\ (\ref{4.2}) and (\ref{4.3})) this leads
to a much higher percolation threshold: while spheres of radius $r=1$ fm
percolate at $n_p^{\rm sphere} \simeq 0.08$ fm$^{-3}$, the threshold
for strings of that length and of radius 0.25 fm (taken as corresponding
ellipsoids) becomes more than ten times higher, with $n_p^{\rm string}
\simeq 1.06$ fm$^{-3}$. For (non-Goldstone) hadrons having a generic mass
of about 1 GeV, this leads to deconfinement at an energy density around
1 GeV/fm$^3$, in general agreement with lattice results.
Equating Eqs.\ (\ref{4.2}) and (\ref{4.3})
gives
\be
r^{\rm core} \simeq 0.46~{\rm fm} \label{5.2}
\ee
as the radius for the percolation of spheres equivalent to the given
string percolation. The `hard core' needed to define deconfinement 
thus has a radius of about half that of a hadron.

\medskip

Once the problem of extending the Coniglio-Klein cluster definition to
the continuum is solved, we can propose to identify the onset of
deconfinement with the percolation threshold for strings of radius $r$ 
and hadronic length $l$. In the limit of infinite quark mass, the critical
behaviour of $SU(N)$ gauge theory and the corresponding percolation
then coincide by virtue of the Coniglio-Klein construction.
For any finite value of the bare quark mass, the QCD counterpart
of the Kert\'esz line will separate a high density colour conducting
deconfined region -- the percolation regime -- from a low density
confined colour insulator, the finite cluster regime.
In the confined region, the average Polyakov loop does not vanish; it
is only exponentially small, $L \sim \exp(-M_h/T)$, because of occasional
thermal string breaking. On the other hand, the average cluster size in
this region is finite, so that the percolation probability remains zero
up to the percolation threshold.

\bigskip

\noindent
{\bf 6.\ Deconfinement in Nuclear Collisions}

\medskip

While deconfinement in an equilibrated QCD medium thus should start at
the percolation threshold for a system of randomly oriented strings,
the pre-equilibrium situtation encountered in nuclear
collisions provides an alignment or polarization of these strings.
Consider a central high energy collision of two identical heavy nuclei,
in which essentially all nucleons undergo several interactions.
These occur in such rapid succession that the nucleons cannot `recover'
before being hit again.
Each individual nucleon-nucleon collision establishes a colour flux
tube or string between the collision partners; the additive quark model
\cite{quarkmodel} suggests that at present (SPS) energies it connects
two triplet colour charges, so that it this just the string considered
above \cite{Luescher,Laermann}. For nuclear collisions in their early
stage, we thus obtain a spaghetti-like structure of intertwined,
overlapping QCD strings, and a cut in the transverse plane results in
the picture shown in Fig.\ 5, giving a distribution of transverse
string areas over a circle of nuclear area. Hence the two-dimensional
continuum percolation theory presented above is the tool to study what
happens when these strings overlap more and more to form connected
spatial regions of increasing size in the transverse plane\footnote{The
first study of this kind was carried out in a specific phenomenological
string model \cite{Pajares}.}. One important modification is that the
overall size of the transverse area is now fixed by nuclear dimensions,
requiring the study of finite size cluster formation.

\medskip

We first consider the cluster density in the transverse plane,
then the geometric cluster size as function of the density of the
overall density of string discs. To have a specific case, we set $R/r
= 20$, corresponding to $R=5$ fm and $r=0.25$ fm; to eliminate some
of the dependence on the specific choice of parameters, it is
convenient to make all quantities ($n,~\n$ and $\S$) dimensionless by
defining them in terms of $\pi r^2$.
In Fig.\ 6 the resulting cluster density \N~is
seen to increase monotonically with the overall density $n$.
As expected from percolation theory, the size $\S$ of the
cluster, in contrast to its density, shows a dramatic variation with
$n$. As noted in section 4, $\S$ diverges for $R/r  \to \infty$
at the percolation point $n_p=1.175$, i.e., when the overall density is
somewhat above one disc per disc area. Of course
there is a considerable overlap of discs, and at $n_p$, the ratio of
the area covered by discs to the overall
area becomes $1-\exp[-1.12] \simeq 0.67$; it approaches unity only
for $n \to \infty$. For finite $R/r $, the critical behaviour is
softened by finite size effects, and for the value $R/r =20$ chosen
above, we obtain the cluster size behaviour shown in Fig.\ 7.
It indeed shows the strongest variation around the percolation point.

\begin{figure}[h]
\vspace*{-10mm}
\centerline{\psfig{file=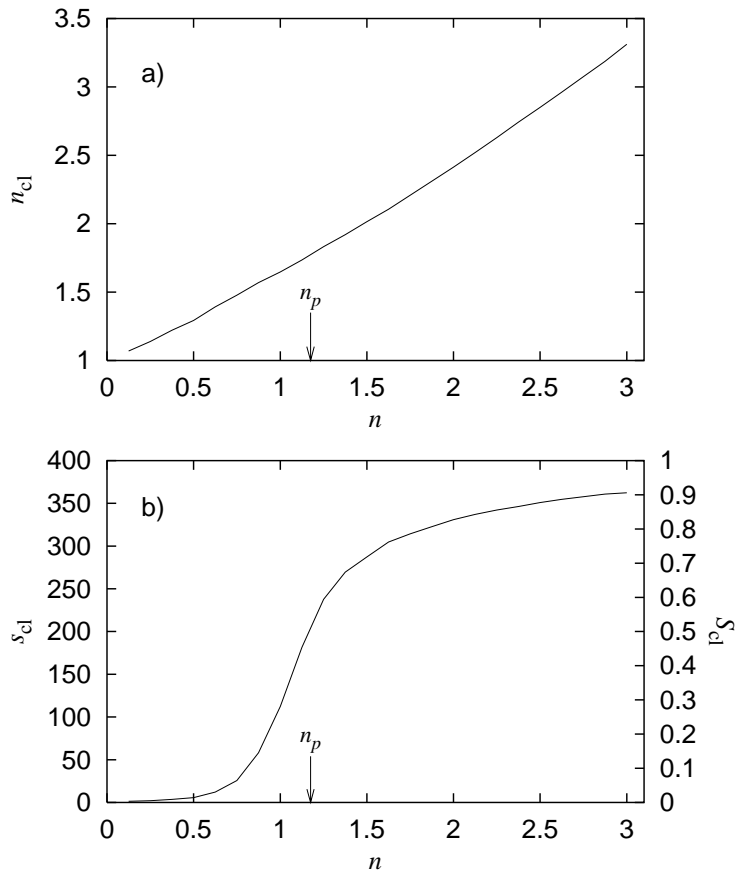,height= 140mm,angle= 0}}
\vspace*{-10mm}
\caption{Cluster density (a) and cluster size (b) as function of overall 
density, in two dimensions, with $r=0.25$ fm, $R=5$ fm; $n_p$ denotes 
the infinite area percolation point.}
\end{figure}

\medskip

Deconfinement is expected to set in when there is
enough `internetting' between interacting nucleons \cite{KLNS}, i.e.,
when the collision density within a cluster becomes sufficiently high;
we denote this value by $\n^c$. It is reached at a certain value
$n_c=n(\n^c)$ of the overall density, which can, but does not need
to be the ultimate percolation point. By Fig.\ 7, this identifies
a corresponding critical cluster size $\S^c = \S(\n^c)$. In other
words, requiring a critical collision density automatically forces the
cluster to have a certain minimum size. This result answers the
question of how much medium is needed before one can speak about a new
phase: systems of geometric size $\S < \S^c$ will on the average not
have a string density sufficient for deconfinement.

\medskip

The onset of deconfinement in nuclear collisions is thus governed by
two parameters: the deconfinement string density $\n^c$ (which could be
the percolation value) and the transverse string size $r$ (which is
roughly determined by QCD \cite{Luescher,Laermann}). The size of
the nuclear transverse area $\pi R^2$ is determined by nuclear
geometry. With $\n^c$ and $r$ fixed, we can then determine the size
$\S^c$ of the bubbles of deconfined medium present at the deconfinement
threshold. Realistic nuclear distributions
\cite{W-S} and a Glauber-based formalism must be used to calculate the
distribution of collisions in the transverse plane as function of the
impact parameter in $A-B$ collisions \cite{KLNS}. The distribution of
the strings in the transverse plane is then no longer random, but
governed by the nucleon-nucleon collision density thus determined.
The resulting formalism can be used to study the onset of deconfinement,
using the predicted \J~suppression as
indicative signal \cite{Matsui}. Here the finite cluster size at the
deconfinement threshold leads to an abrupt onset of such suppression
\cite{NS}, which agrees well with the recently reported anomalous
\J~suppression in $Pb-Pb$ collisions at the CERN-SPS \cite{NA50}.
Moreover, the critical density $n_c$ needed here also agrees rather well with
the (infinite volume) percolation point.

\bigskip

\centerline{\bf Acknowledgements}

\medskip

It is a pleasure to thank Ph.\ Blanchard, F.\ Karsch, M.\ Nardi
and D.\ Stauffer for helpful comments and discussions.

\end{document}